## Main Manuscript for

Nanofluidic logic based on chiral skyrmion flows

Xichao Zhang[a], Jing Xia[b], Yan Zhou[c], Guoping Zhao[b], Xiaoxi Liu[d],
Yongbing Xu[e, f, g, h, *], Masahito Mochizuki[a, *]

[a] Department of Applied Physics, Waseda University, Okubo, Shinjuku-ku, Tokyo 169-8555, Japan

[b] College of Physics and Electronic Engineering, Sichuan Normal University, Chengdu 610068, China

[c] Guangdong Basic Research Center of Excellence for Aggregate Science, School of Science and Engineering, The Chinese University of Hong Kong, Shenzhen, Guangdong 518172, China

[d] Department of Electrical and Computer Engineering, Shinshu University, 4-17-1 Wakasato, Nagano 380-8553, Japan

[e] National Key Laboratory of Spintronics, Nanjing University, Suzhou 215163, China

[f] Jiangsu Provincial Key Laboratory of Advanced Photonic and Electronic Materials, School of Electronic Science and Engineering, Nanjing University, Nanjing 210093, China

[g] School of Integrated Circuits, Nanjing University, Suzhou 215163, China

[h] York-Nanjing International Center for Spintronics (YNICS), School of Physics, Engineering and Technology, University of York, York, YO10 5DD UK

**\*Corresponding Authors**: Yongbing Xu, Masahito Mochizuki
**Emails:** ybxu@nju.edu.cn, masa_mochizuki@waseda.jp

**This PDF file includes:**
    Main Text
    Figures 1 to 5

**Abstract**

Particle-like chiral magnetic skyrmions can flow in nanotracks and behave like chiral fluids. Using interacting flows to perform logical operations is an important topic in microfluidics and nanofluidics. Here, we report a basic nanofluidic logic computing system based on chiral magnetic skyrmions flowing in parallel pipelines connected by an H-shaped junction. The flow behaviors could be manipulated by adjusting the spin polarization angle, which controls the intrinsic skyrmion Hall angle. We demonstrate that, within certain range of the spin polarization angle, fully developed skyrmion flows could lead to fluidic logical operations, which significantly reduce the complexity of skyrmion logic as there is no need for deterministic creation, precise control, and detection of a single isolated skyrmion. Our results suggest that the chiral flow behaviors of magnetic quasiparticles may offer possibilities for spintronic and nanofluidic functions.

**Significance Statement**

Magnetic skyrmions are particle-like topological spin textures in chiral magnets, which can be used as information carriers in spintronic devices. Our work is a piece of research to apply nanofluidic science to skyrmion-based functional applications, leveraging both the particle-like and fluid-like properties of chiral magnetic skyrmions. Our findings suggest that magnetic skyrmions can be functioned as chiral fluids and one can realize skyrmion-based logic computing in a simplified but more robust fluid manner.

**Main Text**

**Introduction**

The world is full of materials and substances that can flow and deform like fluids (**1**). Common fluids in daily life include liquids (**1**) and gases (**2**), while typical fluids also include plasmas (**3**) and granular solid particles (**4**). In principle, all fluids and fluid-like materials can flow and demonstrate complex but appealing dynamic behaviors (**1-16**), such as laminar and turbulent dynamics (**5-10**). The manipulation and control of fluids in artificial structures without mechanical moving parts could lead to promising practical applications, such as logical operations, which are the focus of the field of fluidics (**11-16**). It is therefore important to study all kinds of flow and transport phenomena as they may play important roles in industrial applications and daily human activities.

Skyrmions are topologically non-trivial localized field configurations that can arise in various physical systems, such as liquid crystals (**17-21**) and magnets (**22-36**). In magnetic materials, the chiral exchange interactions could give rise to the presence of chiral skyrmions (**22-35**), which can be treated as nanoscale particles (**36**). A single isolated skyrmion could show topology-dependent particle dynamics (**22-36**) and a large number of particle-like skyrmions may also demonstrate certain fluid-like behaviors (**37, 38**). In particular, the transport of skyrmions in artificial geometries may generate interesting flow behaviors, such as laminar and disordered flows (**37, 38**). The non-trivial topology of skyrmions may also generate exotic phenomena that cannot be reproduced by conventional Newtonian fluid flows (**1, 36-38**), although the skyrmion flow is actually a flow of spin texture patterns (similar to water surface waves), which is not a flow of real fluid particles like water molecules. The Newtonian fluid particles move in the direction of the driving force (e.g., the gravity), while the particle-like skyrmions may move in the direction perpendicular to the driving force due to its topological nature (i.e., the skyrmion Hall effect) (**39-41**). The peculiar properties of skyrmions compared to conventional fluids could lead to novel transport phenomena, which are vital for functional devices (**42-58**).

Magnetic materials are essential components for electronic and spintronic devices that perform data storage and logic computing (**44-47**). An innovative and practical device is the racetrack-type memory based on magnetic domain walls (**59, 60**) or skyrmions (**42, 48**). In 2015, Zhang et al. (**43**) proposed the use of skyrmions in magnetic logic gates based on racetrack-type structures, where the AND and OR gates could be realized based on the conversion between skyrmions and domain walls in a Y-shaped junction. The width of the junction track is close to the skyrmion diameter so that a single skyrmion could be transformed into a pair of domain walls for the purposes of duplication and merging through a fan-out structure. Since then, many studies have been carried out to design logic gates and circuits based on the manipulation of single or multiple isolated skyrmions (**43, 46, 49-54**), which contribute significantly to the emerging field of *skyrmion-based electronics*.

However, the common designs of skyrmion logic are usually based on deterministic creation, precise manipulation, and detection of single or multiple isolated skyrmions (**43, 46, 49-54**), which are challenging tasks especially when the system dimensions are down to nanoscale. The concepts in microfluidics and nanofluidics offer the possibility to design a more robust logic computing system that does not depend on the dynamics of a single nanoscale skyrmion (**11-16**), where logical operations are performed by controlling fluid flows instead of single isolated fluid particles (**11-16**). In this work, we demonstrate computationally that some basic nanofluidic logic could be realized based on many chiral magnetic skyrmions flowing through an H-shaped junction.

## Results and Discussion

**Device Concept and Simulation.** The concept and structure of our proposed device is illustrated in **Fig. 1**. The fluidic logical operations are based on stable flows of particle-like skyrmions flowing through an H-shaped junction (**Fig. 1***A*). The H-shaped junction is made of a ferromagnetic layer and has four ports for the transport of skyrmions, where the left ports are connected to input chambers via parallel pipeline-like nanotracks.

The input chambers (i.e., reservoirs) are used as a source of input skyrmions, where skyrmions are created and stored in a dynamic fashion. It is unnecessary to create and control skyrmions in the chamber in a precise and deterministic manner, however, the number of skyrmions stored in the chamber should be monitored. In principle, the skyrmions in the chamber could be created by using magnetic tunnel junction (MTJ) devices or by using notches (**55-57**). Other current-driven methods could also be used to create skyrmions in the chamber (**58**). Once the two input chambers have stored many skyrmions, it will be possible to carry out logical operations.

To implement the logical operation, one needs to apply a driving force in either one or both two input chambers, which drives skyrmions from the chambers into the H-shaped junction through pipeline nanotracks (i.e., Input A and Input B in **Fig. 1***B*). The driving force is provided by the damping-like spin-orbit torque generated by the spin Hall effect in the heavy-metal substrate layer underneath the ferromagnetic layer (**37, 40, 61, 62**). When a large number of skyrmions are driven into the H-shaped junction in the form of one or two stable flows, they may then flow into either one or both output nanotracks at the right side of the device (i.e., Output A and Output B in **Fig. 1***B*), which depends on the interactions between the skyrmion flows and the H-shaped junction. The interactions can be controlled by geometric and driving parameters, such as the preset driving current direction. Finally, the skyrmions flowing into the output nanotracks will be sent to the chambers at the right side of the device, where they may be stored and reused for cascading computing. It is noteworthy that the interactions between individual skyrmions are of repulsive nature, and the spacing between adjacent skyrmions in a skyrmion flow could be reduced in the presence of strong inter-skyrmion compression (see *SI Appendix*, Supporting Information Text). This feature makes the skyrmion flow more like a compressible flow (like air flow), which is different from incompressible fluids, of which the material density cannot vary over time, such as water.

The input and output electric signals could be detected by four MTJ readout sensors placed upon the two input and two output parallel nanotracks (**Fig. 1***A*). We note that the MTJ-based detection of skyrmion flows in the proposed device differs from the detection of a single isolated skyrmion, which does not require precise control of the skyrmion position under the MTJ sensor. When the skyrmions are flowing through the MTJ readout region, the readout sensor will generate a dynamic electric signal that reflects the number of skyrmions inside the readout region, which is like the situation in skyrmion-based artificial synapses (**55-57**). The digital binary signal is therefore determined by a threshold number of skyrmions inside the MTJ readout region.

In the simulation, we focus on the skyrmion flow dynamics within the H-shaped junction area (**Fig. 1***B*). Specifically, we consider an H-shaped junction area with an overall length of 1000 nm and an overall width of 500 nm. The thickness of the ferromagnetic layer equals 1 nm. The skyrmion nanotracks are formed by locally enhancing the magnetic anisotropy in the ferromagnetic layer (see **Materials and Methods**), where black regions with enhanced anisotropy form barriers that confine and protect skyrmions in the pink nanotracks (**Fig. 1B**). The widths of all pipeline nanotracks are equal to 175 nm. The channel connecting the upper and lower pipeline nanotracks has a length of 500 nm and a width of 100 nm. We note that the geometric parameters of the device, including the size and dimension, are adjustable in principle, but should allow transport of many skyrmions. The skyrmion dynamics is simulated based on the time-dependent Landau-Lifshitz-Gilbert (LLG) equation augmented with the damping-like spin torque generated by the spin Hall effect (see **Materials and Methods**) (**37, 40, 61, 62**).

**Nanofluidic Logic Based on Skyrmion Flows.** In **Fig. 2**, we show possible fluidic logical operations based on stable skyrmion flows in the H-shaped junction. As the H-shaped junction has four ports, we define the left two as the input ports (i.e., Input A and Input B) and the right two as one output port and one vent port.

We assume that the presence of many skyrmions under the MTJ readout sensor (**Fig. 1A**), either in static or dynamic way, will be detected in real-time and lead to the binary signal "1" or "0". As the number of skyrmions in the MTJ readout region depends on the skyrmion flow state, such as the flow speed and the density of skyrmions, the value of the binary signal will be determined by a threshold number of skyrmions detected by the MTJ sensor. For example, if the maximum number of skyrmions that can fill up the MTJ readout region is equal to $N$, then the threshold number for the determination of the binary signal could be defined as $N/2$. Therefore, for both input and output ports, a binary signal of "1" will be obtained when the detected skyrmion number is beyond $N/2$, otherwise, a binary signal of "0" will be obtained.

To implement basic fluidic logic, we need to define the binary output based on the detected state from only one output port. As shown in **Fig. 2A**, the output result is given by the state detected in the upper output port, where the lower output port is used as a vent port. Also, the output result may be given by the state detected in the lower output port (**Fig. 2B**), where the upper output port is used as a vent port.

In **Fig. 2C**, we illustrate five types of possible skyrmion flow pathline patterns in the H-shaped junction. The patterns depend on the skyrmion flow behaviors and the interactions among skyrmions. In the type-I pattern, a stable flow of skyrmions with or without the skyrmion Hall effect are injected into the H-shaped junction from Input A or Input B, where all skyrmions only flow in the upper or lower pipeline nanotrack. Hence, the two input port states are opposite and the two output port states are opposite. A similar case is the type-V pattern, where two stable flows of skyrmions with or without the skyrmion Hall effect are injected, which fill up all nanotracks. Therefore, all input and output port states are the same; with all binary values being "1".

In the type-II pattern, a stable flow of skyrmions with strong skyrmion Hall effect are injected from Input A or Input B, where all skyrmions deflect at the channel connecting the two parallel nanotracks and flow from the upper(lower) nanotrack to the lower(upper) nanotrack. In such a situation, the two input port states are opposite and the two output port states are opposite.

In the type-III pattern, a stable flow of skyrmions with moderate skyrmion Hall effect are injected from Input A or Input B, which is bifurcated into two flows. Hence, the two input port states are opposite, while the two output port states are identical. The output port states will be "1" provided that the bifurcated skyrmion flows can fill up two output nanotracks. Similarly, in the type-IV pattern, two stable flows of skyrmions with strong skyrmion Hall effect are injected from Input A and Input B, which could be merged into a single flow. Therefore, both two input port states are with the binary value "1", while only one output port state will be "1".

In **Fig. 2D**, we show the truth table and corresponding flow behaviors for the H-shaped junction functioned as an OR gate, that is, Output = Input A + Input B. In **Fig. 2E**, we show the truth table and corresponding flow behaviors for the H-shaped junction functioned as an AND gate, that is, Output = Input A · Input B. For the two logic gates, the skyrmion flow behaviors in the H-shaped junction are identical, however, the logic function depends on which MTJ sensor placed on the output ports is working for the detection. The MTJ sensor placed on the vent port could be deactivated.

**Intrinsic Skyrmion Hall Angle and Driving Method.** To implement possible logical operations, it is necessary to control and adjust the intrinsic skyrmion Hall effect, which is an important factor affecting the skyrmion flow behavior in the H-shaped junction. In **Fig. 3A**, we study the intrinsic skyrmion Hall angle $\theta_{SkHE}$ as a function of the spin-polarization angle $\theta$ in the absence of skyrmion-skyrmion and skyrmion-edge interactions, where the in-plane angle between the skyrmion velocity $\boldsymbol{v}$ and the +$x$ direction is defined as $\theta_{SkHE}$ and the in-plane angle between the unit spin polarization vector $\boldsymbol{p}$ (**Eq. 3**) and the +$x$ direction is defined as $\theta$. Namely, $\theta_{SkHE} = \arctan(v_y/v_x)$, where arctan is a function returning the inverse of the tangent function.

The value of $\theta$ could be controlled by adjusting the flow direction of in-plane electrons in the heavy-metal layer according to the spin Hall effect (**40, 61, 62**). In order to change the in-plane electron current direction in experiments, one could fabricate multiple angle-dependent electrodes around the logic gate specimen (**Fig. 3B**).

The current-driven motion of a skyrmion could be described by the Thiele equation (**35, 37, 62, 63**)

$$\boldsymbol{G} \times \boldsymbol{v} - \alpha \mathcal{D} \cdot \boldsymbol{v} - 4\pi \mathcal{B} \cdot \boldsymbol{j}_e = \boldsymbol{0}, \tag{1}$$

where $\boldsymbol{G} = (0,0,-4\pi Q)$ is the gyromagnetic coupling vector associated with the topological charge of skyrmion, that is, $Q = \frac{1}{4\pi} \int \boldsymbol{m} \cdot (\partial_x \boldsymbol{m} \times \partial_y \boldsymbol{m}) \, dx dy$. $\mathcal{D}$ is a dissipative tensor with the diagonal entries being $\mathcal{D}$ and off-diagonal entries being zero. $\mathcal{B}$ is a term quantifying the driving force efficiency. $\boldsymbol{j}_e = (-j_e \sin\theta, j_e \cos\theta)$ is the electron current applied in the heavy-metal substrate. It is worth mentioning that the term $\boldsymbol{G} \times \boldsymbol{v}$ in the left-handed side of **Eq. 1** describes the topological

Magnus force acting on a moving skyrmion, of which the direction is perpendicular to the skyrmion velocity. Such a Magnus force may lead to the skyrmion Hall effect; it is analogous to the Magnus lift force acting on a spinning football in viscous air, of which the path may be deflected (**64**). From **Eq. 1**, one could find that $\theta_{SkHE}$ depends on the value of $\theta$ (**37**), expressed as

$$\theta_{SkHE} = \arctan\left(\frac{\alpha\mathcal{D}\cos\theta + Q\sin\theta}{Q\cos\theta - \alpha\mathcal{D}\sin\theta}\right). \qquad [2]$$

Therefore, the intrinsic skyrmion Hall angle induced by the damping-like spin torque can be linearly adjusted by changing the value of $\theta$. Based on the simulated motion of a single skyrmion and the $\theta_{SkHE}$-$\theta$ relation given in **Eq. 2**, we find $\mathcal{D} = 1.28$ for a skyrmion with $Q$ = -1 relaxed in our model. In the absence of skyrmion-skyrmion and skyrmion-edge interactions, the skyrmion could move toward the right direction with $\theta_{SkHE} \in (-90°, 90°)$ when the spin-polarization angle is set to $\theta \in (111°, 291°)$.

We focus on the skyrmion flow dynamics driven by a small current of $j$ = 0.5 MA cm$^{-2}$ and the spin-polarization angle is within the range between $\theta = 111°$ and $\theta = 291°$, which ensures that the skyrmions are intrinsically moving from input ports to output and vent ports. The current-driven skyrmions may be deformed and their Hall angle may be changed when they are interacting with each other and device edges. As the skyrmion-skyrmion and skyrmion-edge interactions also depend on the skyrmion flow speed (**37**), the deformation may also happen when the driving current is strong. In this work, we focus on the skyrmion flow dynamics without obvious skyrmion deformation, where the skyrmion dynamics is manipulated by $\theta$ instead of $j$.

In **Fig. 3C**, we put 158 skyrmions with $Q$ = -1 in the H-shaped junction, which fill up all nanotracks and are fully relaxed before the application of driving current. The relaxed skyrmions form a triangular lattice (**Fig. 3D**). We also apply the periodic boundary condition in the $x$ direction, so that we can study fully developed skyrmion flow dynamics without including the skyrmion chambers in the model to supply skyrmions, which significantly simplifies the simulation and enhances computational efficiency. With periodic boundary conditions applied in the $x$ direction, the simulated model is in principle equivalent to a model with series connection, where the output ports of the H-shaped junction are connected to the input ports of an adjacent one. Therefore, for certain cases, the simulated model could demonstrate possible skyrmion flow behaviors in both isolated and connected junctions (see **Materials and Methods**).

**Many Skyrmions Flowing Through an H-shaped Junction.** In **Fig. S1** (*SI Appendix*), we show the overlay of flowing skyrmions during the initial stage of simulation for selected values of $\theta$. The skyrmions pathlines at the device center are basically in line with their intrinsic flow direction determined by $\theta_{SkHE}$ in **Eq. 2**, except those close to the edges, where repulsive skyrmion-skyrmion and skyrmion-edge interactions could significantly affect the skyrmion flow direction. It demonstrates that the skyrmion flow direction can be effectively manipulated by varying $\theta$, giving rise to the preference of a skyrmion flow direction toward either one output or both output and vent ports.

For example, when $\theta$ is set to 201 degrees, the intrinsic skyrmion Hall angle equals zero, and therefore, all skyrmions in the device flow from two input ports toward output and vent ports, which is confirmed by the pathlines obtained for the full simulation time frame (**Fig. 4**) as well as the pathlines of fully developed skyrmion flows at the final stage of simulation (see *SI Appendix*, **Fig. S2**). For such a case, the skyrmion flows are able to form type-I and type-V flow behaviors (**Fig. 2C**).

When $\theta$ is set to 251 degrees, the intrinsic flow direction of skyrmions is significantly toward the upper pipeline nanotrack due to strong intrinsic skyrmion Hall effect (see *SI Appendix*, **Fig. S1**). Hence, in principle the skyrmions from either input port will flow into the upper output port, which can be seen from the pathlines of fully developed skyrmion flow in **Fig. S2** (*SI Appendix*). However, if there is a stable supply of skyrmions from both two input ports, the skyrmion flows will be able to fill up all nanotracks, which can be seen from the pathlines given in **Fig. 4**. For such a case, the skyrmion flows could form type-I, type-II, type-IV, and type-V flow behaviors (**Fig. 2C**), depending on which input port is active as well as the supply rate of skyrmions from the input chambers. Particularly, the difference between the supply rate of the two input chambers could determine the transition between type-IV and type-V flow behaviors.

When $\theta$ is set to 211 degrees, the intrinsic skyrmion flow direction is slightly toward the upper pipeline nanotrack due to moderate intrinsic skyrmion Hall effect (see *SI Appendix*, **Fig. S1**). Therefore, the skyrmions injected from Input B are able to flow into both output and vent ports (**Fig. 4** and **Fig. S2** in *SI Appendix*), which could lead to the type-III flow behavior (**Fig. 2C**).

It can be seen from the above three cases that the five types of flow behaviors in **Fig. 2C** can be realized based on the manipulation of $\theta$, which justifies that the H-shaped junction could be used to carry out some basic fluidic logical operations. However, the skyrmion flow dynamics may be strongly affected by the skyrmion-skyrmion and skyrmion-edge interactions. For example, when $\theta$ is set to 181 degrees, the skyrmion-skyrmion and skyrmion-edge repulsions result in a static accumulation of skyrmions near the lower pipeline nanotrack (see *SI Appendix*, **Fig. S2**) and when $\theta$ is set to 131-171 degrees, the skyrmions in the fully developed stage flow toward the input ports (**Fig. 4** and **Fig. S2** in *SI Appendix*).

Based on the above results, we demonstrated in **Fig. 5** that fully developed skyrmion flows can be used to realize logical OR and AND operations given in **Fig. 2D** and **Fig. 2E**, respectively, where the spin polarization angle is fixed at a suitable value, such as $\theta$ = 231 degrees (see *SI Appendix*, **Movie S1-S3**).

It is worth noting that although many skyrmions flowing through pipeline nanotracks and H-shaped junctions could behave like fluid particles and exhibit certain fluid-like behaviors at low Reynolds numbers, such as laminar flow, bifurcation, and convergence, there are some notable differences between particle-like skyrmions and conventional fluid particles. For example, magnetic skyrmions are nanoscale quasiparticles formed by localized spin textures in magnetic materials, while fluids are made of atoms or molecules that are able to move around one another. The motion of skyrmions is induced by spin torques, while the fluid motion is driven by external forces, such as pressure gradients and gravity. Also, there is no momentum conservation in a steady laminar flow of massless skyrmions. In principle, the mass of skyrmions could be induced when skyrmions are dynamically interacting with each other or with magnons (**65**).

**Advantages and Disadvantages of Nanofluidic Logic Based on Skyrmion Flows.** Skyrmion-based logic computing has been extensively studied for ten years due to its importance in future spintronic circuits. The most remarkable advantage of using isolated skyrmions as information elements is that the nanoscale size of a single skyrmion could result in very high device bit-storage density and device packing density. Another advantage of skyrmions is their great potential for building devices with low energy consumption. Therefore, basically all skyrmion-based logic gates proposed in the past decade are based on the manipulation of single or multiple isolated skyrmions (**43, 46, 49-54**).

However, logic gates based on isolated skyrmions may face several challenges and difficulties, particularly in the precise and deterministic creation, position control, and detection of a single skyrmion. For example, to carry out logical operations based on isolated skyrmions, the device needs to create isolated skyrmions at the input ports in a precise manner, and then, the device needs to control the position of a single skyrmion or the spacing between two shifting skyrmions in the channel connecting the input and output ports. At the output ports the device needs to readout a single nanoscale skyrmion using an MTJ sensor or other electric methods. All these processes are difficult to implement at the nanoscale due to factors such as pinning effects and fluctuations in the device.

To overcome these difficulties and facilitate the experimental realization of skyrmion-based logic gates, fluidic logic gates based on skyrmion flows could be a compromise solution, which provides greater feasibility and flexibility for skyrmion-based logic computing. In fluidic logic gates, the information is carried by skyrmion flows instead of single isolated skyrmions. For this reason, it is unnecessary to fabricate nanotracks with dimensions close to or lower than the skyrmion size.

The duplication and merging of information carriers can be performed by the bifurcation and convergence of skyrmion flows, which do not require the destruction and recreation of skyrmions or the transformation between skyrmions and domain walls. And more importantly, in fluidic logic gates, it is unnecessary to fabricate complex notches and pinning sites to control and manipulate a single skyrmion. The spacing among skyrmions may be changed in real devices, however, it will not affect the skyrmion flow and the information carried by the flow, provided that the flow is fully developed.

At the input ports of fluidic logic gates, many skyrmions are created and stored in chambers, and then delivered to the computing element, creating a dynamic supply of skyrmions. At the output ports, the detection of many skyrmions using MTJ sensors would be an easier task, as it eliminates the requirement to fabricate MTJ sensors and nanotracks down to the dimension of a single skyrmion (i.e., a few nanometers).

The thermal annihilation of skyrmions will cease to be a major concern in fluidic logic gates as the dynamic supply of skyrmions from the input chamber could in principle ensure fully developed skyrmion flows. The dynamic supply of skyrmions could also support a dynamic computing architecture, where the system experiences no delay between the input and output logic ports. As there is no need to destroy and recreate skyrmions during fluidic logic computing, it is possible to replace skyrmions by other particle-like topological spin textures, such as bimerons (**66**).

On the other hand, we would like to point out that the main drawback of using skyrmion flows for computing, and even for storage, is the need to fabricate larger chambers and wider nanotracks compared to those used for isolated skyrmions. Consequently, the skyrmion-based fluidic devices could be bigger than single-skyrmion-based devices, which may lead to a reduction in the device packing density. Besides, as fluidic logic gates based on skyrmions need to ensure that many skyrmions in the device form steady laminar (low-Reynolds-number) flows, it may be difficult to operate skyrmions at ultra-high speeds. However, in principle, the energy consumption of current-driven skyrmions does not depend on the number of skyrmions, as the driving current is usually applied to the entire ferromagnetic layer hosting skyrmions. Therefore, the energy consumption of fluidic logic gates based on skyrmion flows should be comparable to that of single-skyrmion-based devices.

**Conclusion**

In conclusion, we have proposed a transformative approach for realizing basic logic computing based on magnetic chiral skyrmions, which utilizes the flow behaviors of many skyrmions in an H-shaped junction. We computationally demonstrated that the current-driven flow behaviors of skyrmions in the H-shaped junction could be effectively controlled by adjusting the spin-polarization angle, which affects the intrinsic skyrmion Hall angle and, consequently, the interactions between skyrmion flows and the H-shaped junction. We show that several types of skyrmion flow pathlines in the H-shaped junction could be utilized to construct basic logic gates, such as the OR and AND gates. The size of nanofluidic logic gates based on chiral skyrmion flows could be larger than conventional devices based on single isolated skyrmions, however, a modest increase in device size may lead to significant improvements and advancements in skyrmionics technology. In principle, larger devices are easier to fabricate and implement in labs and fabs. The nanofluidic logic gates based chiral skyrmion flows are potentially technologically useful as they can overcome several critical challenges inherent in conventional single-skyrmion-based logic device designs, such as the requirements of deterministic creation, precise control, and detection of a single isolated skyrmion. Nanofluidic devices based on skyrmion flows are also especially useful for information processing under extreme conditions, such as exposure to radiation. Our results uncover the rich and complex flow behaviors of particle-like topological spin textures, and meanwhile, may motivate the development of an interdisciplinary fluid research area focusing on the flow dynamics and applications of chiral quasiparticles in spintronic systems.

**Materials and Methods**

**Computational Simulations.** The dynamics of spins in the ferromagnetic layer were simulated by using the micromagnetic simulator MUMAX$^3$ on several NVIDIA GeForce RTX 3060 Ti graphics processing units (**67**). The time-dependent LLG equation augmented with the damping-like spin torque is given as

$$\partial_t \boldsymbol{m} = -\gamma_0 \boldsymbol{m} \times \boldsymbol{h}_{\text{eff}} + \alpha(\boldsymbol{m} \times \partial_t \boldsymbol{m}) + u(\boldsymbol{m} \times \boldsymbol{p} \times \boldsymbol{m}),$$ [3]

where $\boldsymbol{m}$ is the reduced magnetization, $t$ is the time, $\gamma_0$ is the absolute gyromagnetic ratio, and $\alpha$ is the Gilbert damping parameter. The spin torque coefficient is given by $u = |(\gamma_0 \hbar / \mu_0 e)| \cdot (j\theta_{\text{SH}}/2aM_{\text{S}})$. $\mu_0$ and $M_{\text{S}}$ denote the vacuum permeability constant and the saturation magnetization, respectively. $\hbar$ is the reduced Planck constant, $e$ is the electron charge, $j$ is the applied current density, $a$ is the thickness of the ferromagnetic layer, and $\theta_{\text{SH}}$ is the spin Hall angle. $\boldsymbol{p}$ denotes the spin polarization direction, which can be controlled by varying the electric current direction in the heavy metal substrate. The effective field $\boldsymbol{h}_{\text{eff}} = -\frac{1}{\mu_0 M_{\text{S}}} \cdot \frac{\delta\varepsilon}{\delta\boldsymbol{m}}$, where $\varepsilon$ denotes energy density. The micromagnetic energy density terms considered in the simulation include the ferromagnetic exchange interaction, interface-induced Dzyaloshinskii-Moriya (DM) exchange interaction, perpendicular magnetic anisotropy, and demagnetization terms, which are expressed as follows

$$\varepsilon = A(\nabla\boldsymbol{m})^2 + D[m_z(\boldsymbol{m} \cdot \nabla) - (\nabla \cdot \boldsymbol{m})m_z] - K(\boldsymbol{n} \cdot \boldsymbol{m})^2 - \frac{M_{\text{S}}}{2}(\boldsymbol{m} \cdot \boldsymbol{B}_{\text{d}}),$$ [4]

where $A$ and $D$ stand for the ferromagnetic and DM exchange interaction constants, respectively. $K$ is the perpendicular magnetic anisotropy constant, and $\boldsymbol{B}_{\text{d}}$ is the demagnetization field.

In all simulations, the mesh size is set to $2.5 \times 2.5 \times 1$ nm$^3$ to ensure good computational accuracy and efficiency. Other default parameters are (**37, 42, 43**): $\gamma_0 = 2.211 \times 10^5$ m A$^{-1}$ S$^{-1}$, $\alpha = 0.3$, $M_{\text{S}} = 580$ kA m$^{-1}$, $A = 15$ pJ m$^{-1}$, $D = 3$ mJ m$^{-2}$, and $K = 0.8$ MJ m$^{-3}$. For the sake of simplicity, we assume that $\theta_{\text{SH}} = 1$ and the driving force is controlled by the applied current density straightforwardly. We note that a smaller spin Hall angle will in principle result in a smaller driving force as the spin torque strength $u \sim j\theta_{\mathrm{SH}}$. We also note that the value of $K$ in the regions with enhanced perpendicular magnetic anisotropy is assumed to be ten times larger than the default value, which ensures that the skyrmions driven by the current are confined in the pipeline nanotracks. In experiments, the ferromagnetic layer is usually grown on a heavy-metal substrate layer by using magnetron sputtering. The interface between the ferromagnetic layer and the heavy-metal layer is essential for the generation of DM exchange interaction and perpendicular magnetic anisotropy (**40, 41, 42, 58, 62**), which stabilize chiral skyrmions in the ferromagnetic layer. The spin Hall effect in the heavy-metal layer leads to the generation of spin currents (**59**), which propagate vertically into the ferromagnetic layer and drive the skyrmion flow dynamics. Exemplary material systems are the Co/Pt and CoFeB/Ta bilayers (**40, 42**), where skyrmions in the ferromagnetic Co and CoFeB layers could be driven into motion by spin currents from the heavy-metal Pt and Ta layers (**40, 42**).

It should also be noted that we by default apply the periodic boundary condition in the *x* direction, for example, in simulations for **Fig. 4**, where skyrmions moving out from the right boundary at *x* = 1000 nm will reenter the simulation window from the left boundary at *x* = 0 nm. In such a case, the simulated model is equivalent to a model with series connection, that is, the output ports of the H-shaped junction are connected to the input ports of an adjacent H-shaped junction. Therefore, when skyrmions injected from Input A/B are flowing into Output B/A, they will reenter the simulation window from Input B/A, demonstrating possible skyrmion flow behaviors in connected junctions. If skyrmions injected from Input A/B are flowing into Output A/B, they will reenter the simulation window from Input A/B, which demonstrate possible skyrmion flow behaviors in both isolated and connected junctions. However, in simulations for **Fig. 5**, we removed the periodic boundary condition to show certain flow patterns for logical operations, that is, skyrmions moving out from the right boundary at *x* = 1000 nm are removed from the model and will not reenter the simulation window. In experiments, the skyrmions flowing through the H-shaped junction could be sent to output chambers or simply erased from the device.

**Data, Materials, and Software Availability**

All study data are included in this article and/or SI Appendix. The micromagnetic simulator MUMAX[3] used in this work is publicly accessible at http://mumax.github.io/index.html.

**Acknowledgments**

X.Z. and M.M. acknowledge support by the CREST, the Japan Science and Technology Agency (Grant No. JPMJCR20T1). X.Z. also acknowledges support by the Grants-in-Aid for Scientific Research from JSPS KAKENHI (Grants No. JP25K17939 and No. JP20F20363). M.M. also acknowledges support by the Grants-in-Aid for Scientific Research from JSPS KAKENHI (Grants No. JP25H00611, No. JP24H02231, No. JP23H04522, and No. JP20H00337) and the Waseda University Grant for Special Research Projects (Grant No. 2025C-133). Y.Z. acknowledges support by the Shenzhen Fundamental Research Fund (Grant No. JCYJ20210324120213037), the Guangdong Basic and Applied Basic Research Foundation (Grant No. 2021B1515120047), the Shenzhen Peacock Group Plan (Grant No. KQTD20180413181702403), and the National Natural Science Foundation of China (Grant No. 12374123), Guangdong Basic Research Center of Excellence for Aggregate Science, and the 2023 SZSTI stable support scheme. G.Z. acknowledges support by the State Key Research and Development Program of China (Grant No. 2021YFB3500100) and the Central Government Funds of Guiding Local Scientific and Technological Development for Sichuan Province (Grant No. 2021ZYD0025). X.L. acknowledges support by the Grants-in-Aid for Scientific Research from JSPS KAKENHI (Grants No. JP21H01364, No. JP21K18872, No. JP22F22061, No. JP24K21711, and No. JP25K01606). Y.X. acknowledges support by the National Key Research and Development Program of China (Grants No. 2024YFA1408801, No. 2024YFA1408803, and No. 2021YFB3601600). This work was also supported by the National Natural Science Foundation of China (Grants No. 62427901, No. 12474117, No. 12241403, No. 12374123, No. 12104327, No. 12474122, No. 51771127, No. 52171188, and No. 52111530143).

**References**

1. F. M. White, Fluid Mechanics, 7th Edition. (McGrawHill, New York, 2011).

2. G. Bar-Meir, Fundamentals of Compressible Fluid Mechanics. (Potto Project, 2004).

3. M. Kono, M. M. Škorić, The fluid theory of plasmas, in: nonlinear physics of plasmas, Springer Series on Atomic, Optical, and Plasma Physics, Vol. 62. (Springer, Berlin, Heidelberg, 2010).

4. R. P. Behringer, B. Chakraborty, The physics of jamming for granular materials: A review. *Rep. Prog. Phys.* **82**, 012601 (2019).

5. B. Eckhardt, T. M. Schneider, B. Hof, J. Westerweel, Turbulence transition in pipe flow. *Annu. Rev. Fluid Mech.* **39**, 447-468 (2007).

6. A. Guha, Transport and deposition of particles in turbulent and laminar flow. *Annu. Rev. Fluid Mech.* **40**, 311-341 (2008).

7. T. Zahtila, L. Chan, A. Ooi, J. Philip, Particle transport in a turbulent pipe flow: Direct numerical simulations, phenomenological modelling and physical mechanisms. *J. Fluid Mech.* **957**, A1 (2023).

8. D. Barkley, L. S. Tuckerman, Computational study of turbulent laminar patterns in Couette flow. *Phys. Rev. Lett.* **94**, 014502 (2005).

9. M. C. Miguel, A. Mughal, S. Zapperi, Laminar flow of a sheared vortex crystal: scars in flat geometry. *Phys. Rev. Lett.* **106**, 245501 (2011).

10. J. Knebel, M. F. Weber, E. Frey, In pursuit of turbulence. *Nat. Phys.* **12**, 204-205 (2016).

11. T. M. Squires, S. R. Quake, Microfluidics: Fluid physics at the nanoliter scale. *Rev. Mod. Phys.* **77**, 977 (2005).

12. A. P. Singh, M. Tintelott, E. Moussavi, S. Ingebrandt, R. Leupers, X.-T. Vu, F. Merchant, V. Pachauri, Logic operations in fluidics as foundation for embedded biohybrid computation. *Device* **1**, 100220 (2023).

13. A. P. Antonov, M. Terkel, F. J. Schwarzendahl, C. Rodríguez-Gallo, P. Tierno, H. Löwen, Controlling colloidal flow through a microfluidic Y-junction. arXiv [Preprint] (2024). https://arxiv.org/abs/2406.00883 (accessed 3 June 2024).

14. J. H. Jung, G. Destgeer, J. Park, H. Ahmed, K. Parka, H. Jin Sung, Microfluidic flow switching via localized acoustic streaming controlled by surface acoustic waves. *RSC Adv.* **8**, 3206-3212 (2018).

15. L. Nan, H. Zhang, D. A. Weitz, H. C. Shum, Development and future of droplet microfluidics. *Lab Chip* **24**, 1135-1153 (2024).

16. M. Prakash, N. Gershenfeld, Microfluidic bubble logic. *Science* **315**, 832-835 (2007).

17. A. O. Leonov, I. E. Dragunov, U. K. Rößler, A. N. Bogdanov, Theory of skyrmion states in liquid crystals. *Phys. Rev. E* **90**, 042502 (2014).

18. A. Duzgun, J. V. Selinger, A. Saxena, Comparing skyrmions and merons in chiral liquid crystals and magnets. *Phys. Rev. E* **97**, 062706 (2018).

19. D. Foster, C. Kind, P. J. Ackerman, J.-S. B. Tai, M. R. Dennis, I. I. Smalyukh, Two-dimensional skyrmion bags in liquid crystals and ferromagnets. *Nat. Phys.* **15**, 655-659 (2019).

20. J. S. Wu, I. I. Smalyukh, Hopfions, heliknotons, skyrmions, torons and both abelian and nonabelian vortices in chiral liquid crystals. *Liq. Cryst. Rev.* **10**, 34-68 (2022).

21. T. Alvim, M. M. Telo da Gama, M. Tasinkevych, Collective variable model for the dynamics of liquid crystal skyrmions. *Commun. Phys.* **7**, 2 (2024).

22. A. N. Bogdanov, D. A. Yablonskii, Thermodynamically stable "vortices" in magnetically ordered crystals. The mixed state of magnets. *Sov. Phys. JETP* **68**, 101-103 (1989).

23. U. K. Rößler, A. N. Bogdanov, C. Pfleiderer, Spontaneous skyrmion ground states in magnetic metals. *Nature* **442**, 797-801 (2006).

24. S. Mühlbauer, B. Binz, F. Jonietz, C. Pfleiderer, A. Rosch, A. Neubauer, R. Georgii, P. Böni, Skyrmion lattice in a chiral magnet. *Science* **323**, 915-919 (2009).

25. X. Z. Yu, Y. Onose, N. Kanazawa, J. H. Park, J. H. Han, Y. Matsui, N. Nagaosa, Y. Tokura, Real-space observation of a two-dimensional skyrmion crystal. *Nature* **465**, 901-904 (2010).

26. A. N. Bogdanov, C. Panagopoulos, Physical foundations and basic properties of magnetic skyrmions. *Nat. Rev. Phys.* **2**, 492-498 (2020).

27. N. Nagaosa, Y. Tokura, Topological properties and dynamics of magnetic skyrmions. *Nat. Nanotech.* **8**, 899-911 (2013).

28. R. Wiesendanger, Nanoscale magnetic skyrmions in metallic films and multilayers: a new twist for spintronics. *Nat. Rev. Mat.* **1**, 16044 (2016).

29. G. Finocchio, F. Büttner, R. Tomasello, M. Carpentieri, M. Kläui, Magnetic skyrmions: from fundamental to applications. *J. Phys. D: Appl. Phys.* **49**, 423001 (2016).

30. A. Fert, N. Reyren, V. Cros, Magnetic skyrmions: advances in physics and potential applications. *Nat. Rev. Mater.* **2**, 17031 (2017).

31. W. Jiang, G. Chen, K. Liu, J. Zang, S. G. Velthuiste, A. Hoffmann, Skyrmions in magnetic multilayers. *Phys. Rep.* **704**, 1-49 (2017).

32. K. Everschor-Sitte, J. Masell, R. M. Reeve, M. Kläui, Perspective: Magnetic skyrmions-Overview of recent progress in an active research field. *J. Appl. Phys.* **124**, 240901 (2018).

33. X. Zhang, Y. Zhou, K. M. Song, T.-E. Park, J. Xia, M. Ezawa, X. Liu, W. Zhao, G. Zhao, S. Woo, Skyrmion-electronics: writing, deleting, reading and processing magnetic skyrmions toward spintronic applications. *J. Phys. Condens. Matter* **32**, 143001 (2020).

34. B. Göbel, I. Mertig, O. A. Tretiakov, Beyond skyrmions: Review and perspectives of alternative magnetic quasiparticles. *Phys. Rep.* **895**, 1-28 (2021).

35. Y. Ohki, M. Mochizuki, Fundamental theory of current-induced motion of magnetic skyrmions. *J. Phys.: Condens. Matter* **37**, 023003 (2025).

36. C. Reichhardt, C. J. O. Reichhardt, M. V. Milosevic, Statics and dynamics of skyrmions interacting with disorder and nanostructures. *Rev. Mod. Phys.* **94**, 035005 (2022).

37. X. Zhang, J. Xia, O. A. Tretiakov, M. Ezawa, G. Zhao, Y. Zhou, X. Liu, M. Mochizuki, Laminar and transiently disordered dynamics of magnetic-skyrmion pipe flow. *Phys. Rev. B* **108**, 144428 (2023).

38. K. Raab, M. Schmitt, M. A. Brems, J. Rothörl, F. Kammerbauer, S. Krishnia, M. Kläui, P. Virnau, Skyrmion flow in periodically modulated channels. *Phys. Rev. E* **110**, L042601 (2024).

39. J. Zang, M. Mostovoy, J. H. Han, N. Nagaosa, Dynamics of skyrmion crystals in metallic thin films. *Phys. Rev. Lett.* **107**, 136804 (2011).

40. W. Jiang, X. Zhang, G. Yu, W. Zhang, X. Wang, M. Benjamin Jungfleisch, J. E. Pearson, X. Cheng, O. Heinonen, K. L. Wang, Y. Zhou, A. Hoffmann, S. G. E. Velthuiste, Direct observation of the skyrmion Hall effect. *Nat. Phys.* **13**, 162-169 (2017).

41. K. Litzius, I. Lemesh, B. Kruger, P. Bassirian, L. Caretta, K. Richter, F. Buttner, K. Sato, O. A. Tretiakov, J. Forster, R. M. Reeve, M. Weigand, I. Bykova, H. Stoll, G. Schutz, G. S. D. Beach, M. Kläui, Skyrmion Hall effect revealed by direct time-resolved X-ray microscopy. *Nat. Phys.* **13**, 170-175 (2017).

42. J. Sampaio, V. Cros, S. Rohart, A. Thiaville, A. Fert, Nucleation, stability and current-induced motion of isolated magnetic skyrmions in nanostructures. *Nat. Nanotechnol.* **8**, 839-844 (2013).

43. X. Zhang, M. Ezawa, Y. Zhou, Magnetic skyrmion logic gates: conversion, duplication and merging of skyrmions. *Sci. Rep.* **5**, 9400 (2015).

44. W. Kang, Y. Huang, X. Zhang, Y. Zhou, W. Zhao, Skyrmion-electronics: An overview and outlook. *Proc. IEEE* **104**, 2040-2061 (2016).

45. S. Li, W. Kang, X. Zhang, T. Nie, Y. Zhou, K. L. Wang, W. Zhao, Magnetic skyrmions for unconventional computing. *Mater. Horiz.* **8**, 854-868 (2021).

46. S. Luo, L. You, Skyrmion devices for memory and logic applications. *APL Mater.* **9**, 050901 (2021).

47. C. H. Marrows, K. Zeissler, Perspective on skyrmion spintronics. *Appl. Phys. Lett.* **119**, 250502 (2021).

48. D. Song, W. Wang, S. Zhang, Y. Liu, N. Wang, F. Zheng, M. Tian, R. E. Dunin-Borkowski, J. Zang, H. Du, Steady motion of 80-nm-size skyrmions in a 100-nm-wide track. *Nat. Commun.* **15**, 5614 (2024).

49. S. Luo, M. Song, X. Li, Y. Zhang, J. Hong, X. Yang, X. Zou, N. Xu, L. You, Reconfigurable skyrmion logic gates. *Nano Lett.* **18**, 1180-1184 (2018).

50. M. Chauwin, X. Hu, F. Garcia-Sanchez, N. Betrabet, A. Paler, C. Moutafis, J. S. Friedman, Skyrmion logic system for large-scale reversible computation. *Phys. Rev. Applied* **12**, 064053 (2019).

51. M. G. Mankalale, Z. Zhao, J.-P. Wang, S. S. Sapatnekar, SkyLogic-A proposal for a skyrmion-based logic device. *IEEE Trans. Electron Devices* **66**, 1990-1996 (2019).

52. B. W. Walker, C. Cui, F. Garcia-Sanchez, J. A. C. Incorvia, X. Hu, J. S. Friedman, Skyrmion logic clocked via voltage-controlled magnetic anisotropy. *Appl. Phys. Lett.* **118**, 192404 (2021).

53. X. Liang, J. Xia, X. Zhang, M. Ezawa, O. A. Tretiakov, X. Liu, L. Qiu, G. Zhao, Y. Zhou, Antiferromagnetic skyrmion-based logic gates controlled by electric currents and fields. *Appl. Phys. Lett.* **119**, 062403 (2021).

54. X. He, G. Yu, M. Zhu, Y. Qiu, H. Zhou, Guided motion of a magnetic skyrmion by a voltage-controlled strained channel, and logic applications. *Phys. Rev. Applied* **22**, 054047 (2024).

55. Z. Khodzhaev, J. A. C. Incorvia, Tunable spike-timing-dependent plasticity in magnetic skyrmion manipulation chambers. *Appl. Phys. Lett.* **124**, 262402 (2024).

56. K. M. Song, J.-S. Jeong, B. Pan, X. Zhang, J. Xia, S. Cha, T.-E. Park, K. Kim, S. Finizio, J. Raabe, J. Chang, Y. Zhou, W. Zhao, W. Kang, H. Ju, S. Woo, Skyrmion-based artificial synapses for neuromorphic computing. *Nat. Electron.* **3**, 148-155 (2020).

57. T. da Câmara Santa Clara Gomes, Y. Sassi, D. Sanz-Hernández, S. Krishnia, S. Collin, M.-B. Martin, P. Seneor, V. Cros, J. Grollier, N. Reyren, Neuromorphic weighted sums with magnetic skyrmions. *Nat. Electron.* **8**, 204-214 (2025).

58. W. Jiang, P. Upadhyaya, W. Zhang, G. Yu, M. B. Jungfleisch, F. Y. Fradin, J. E. Pearson, Y. Tserkovnyak, K. L. Wang, O. Heinonen, S. G. E. te Velthuis, A. Hoffmann, Blowing magnetic skyrmion bubbles. *Science* **349**, 283-286 (2015).

59. S. S. P. Parkin, M. Hayashi, L. Thomas, Magnetic domain-wall racetrack memory. *Science* **320**, 190-194 (2008).

60. S. Parkin, S.-H. Yang, Memory on the racetrack. *Nat. Nanotechnol.* **10**, 195-198 (2015).

61. J. Sinova, S. O. Valenzuela, J. Wunderlich, C. H. Back, T. Jungwirth, Spin Hall effects. *Rev. Mod. Phys.* **87**, 1213 (2015).

62. Z. Wang, X. Zhang, J. Xia, L. Zhao, K. Wu, G. Yu, K. L. Wang, X. Liu, S. G. E. te Velthuis, A. Hoffmann, Y. Zhou, W. Jiang, Generation and Hall effect of skyrmions enabled using nonmagnetic point contacts. *Phys. Rev. B* **100**, 184426 (2019).

63. A. A. Thiele, Steady-state motion of magnetic domains. *Phys. Rev. Lett.* **30**, 230 (1973).

64. K. Everschor-Sitte, M. Sitte, Real-space Berry phases: Skyrmion soccer. *J. Appl. Phys.* **115**, 172602 (2014).

65. D. Wang, H.-B. Braun, Y. Zhou, Dynamical mass generation for ferromagnetic skyrmions in two dimensions. *J. Magn. Magn. Mater.* **564**, 170062 (2022).

66. X. Zhang, Y. Zhou, X. Yu, M. Mochizuki, Bimerons create bimerons: proliferation and aggregation induced by current and field. *Aggregate* **5**, e590 (2024).

67. A. Vansteenkiste, J. Leliaert, M. Dvornik, M. Helsen, F. Garcia-Sanchez, B. V. Waeyenberge, The design and verification of MuMax3. *AIP Adv.* **4**, 107133 (2014).

**Figures**

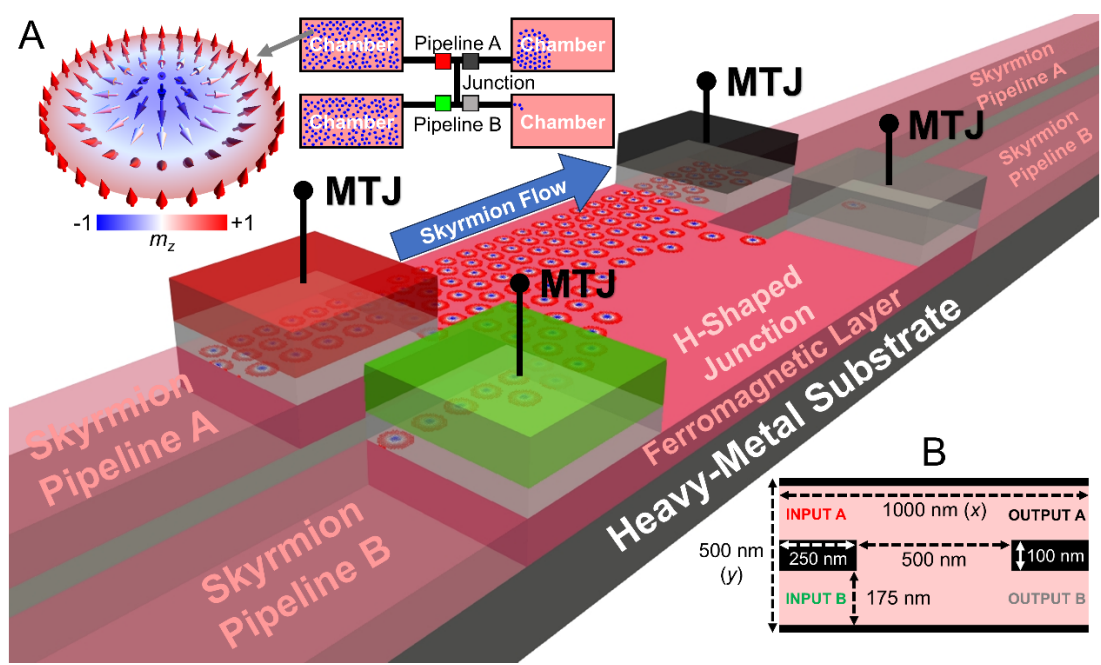

**Fig. 1. Schematic illustration of the nanofluidic logic gate based on many nanoscale skyrmions flowing through an H-shaped junction.** (***A***) Schematic illustration of the spin configuration of a Néel-type chiral skyrmion, which is stabilized by the antisymmetric exchange interaction (i.e., Dzyaloshinskii-Moriya interaction) generated at the ferromagnet/heavy metal interface. The color scale represents the reduced out-of-plane magnetization component $m_z$, which is used throughout the paper. (***B***) Top view of the simulated device geometry, which focuses on the H-shaped computing junction. A number of skyrmions are initially created and stored in input chambers at the left side of the device. They can be driven into the H-shaped junction part through two parallel input pipeline nanotracks (i.e., Input A and Input B). Depending on the interactions among the two flows of skyrmions in the H-shaped junction region as well as the driving parameters (e.g., preset driving current direction), skyrmions may flow into the output and vent pipeline nanotracks. The input and output electric signals could be detected by three MTJ readout sensors placed upon the two input and one output pipelines. The skyrmions used for logic computing will be driven into output and vent chambers at the right side of the device, where they could be reused for cascading computing or erased from the device.

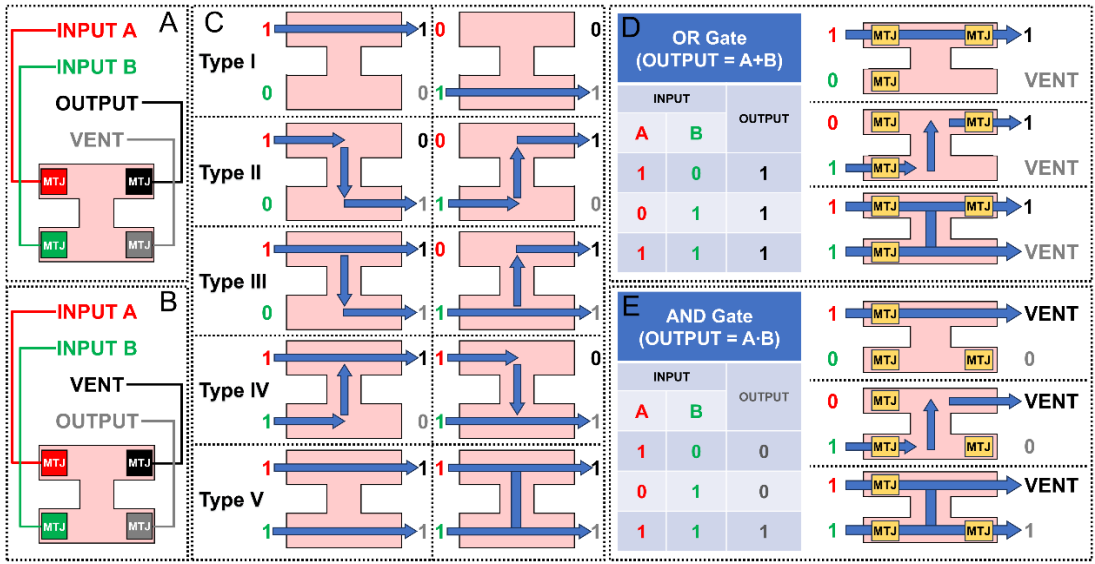

**Fig. 2. Schematics and tables showing possible nanofluidic logic computing schemes and operations based on skyrmion flow behaviors in the H-shaped junction.** (***A***) The logic computation result is given by Output A, while Output B is used as a vent port. (***B***) The logic computation result is given by Output B, while Output A is used as a vent port. (***C***) Possible non-trivial types of skyrmion flow behaviors in the H-shaped junction. Blue arrows indicate the flow directions of the skyrmion flows. The MTJ readout sensors placed upon the input and output pipelines can detect the skyrmions in an electric manner. The readout sensors return logic state "1" in the presence of many skyrmions underneath the sensors, while they return logic state "0" in the absence of skyrmions or in the presence of a few skyrmions underneath the sensors. (***D***) Truth table and corresponding flow behaviors for the H-shaped junction functioned as an OR gate. (***E***) Truth table and corresponding flow behaviors for the H-shaped junction functioned as an AND gate.

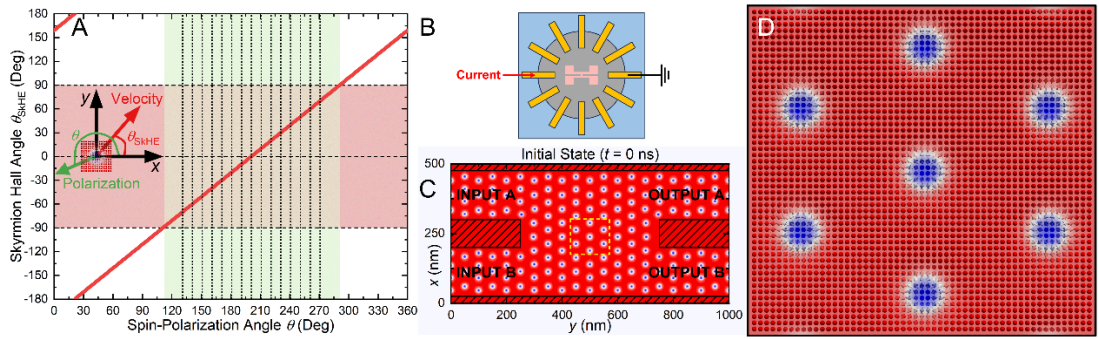

**Fig. 3. Intrinsic skyrmion Hall angle controlled by the driving parameter.** (**A**) Intrinsic skyrmion Hall angle as a function of the spin-polarization angle. To drive the skyrmions flowing from the left input ports to the right output ports through the H-shaped junction, the intrinsic skyrmion Hall angle should be set to the range between -90 and 90 degrees. Hence, the corresponding spin-polarization angle should be adjusted between 111 and 291 degrees. The intrinsic skyrmion Hall angle equals zero when the spin-polarization angle equals 201 degrees. Note that both the intrinsic skyrmion Hall angle and the spin-polarization angle lie in the $x$-$y$ plane. (**B**) Schematic illustration of electrode geometry. The electrodes are indicated by yellow bars. To change the current direction, one may fabricate multiple angle-dependent electrodes around the logic gate specimen. The spin-polarization angle could be adjusted by changing the current direction in the heavy metal. (**C**) Top view of the initial spin configurations in the simulated area of the H-shaped junction. 158 skyrmions with $Q$ = -1 are placed and relaxed in the device, forming a triangular lattice of skyrmions. (**D**) Close-up view of the spin configuration of several skyrmions at the device center, as indicated by the yellow dashed line box in (**C**).

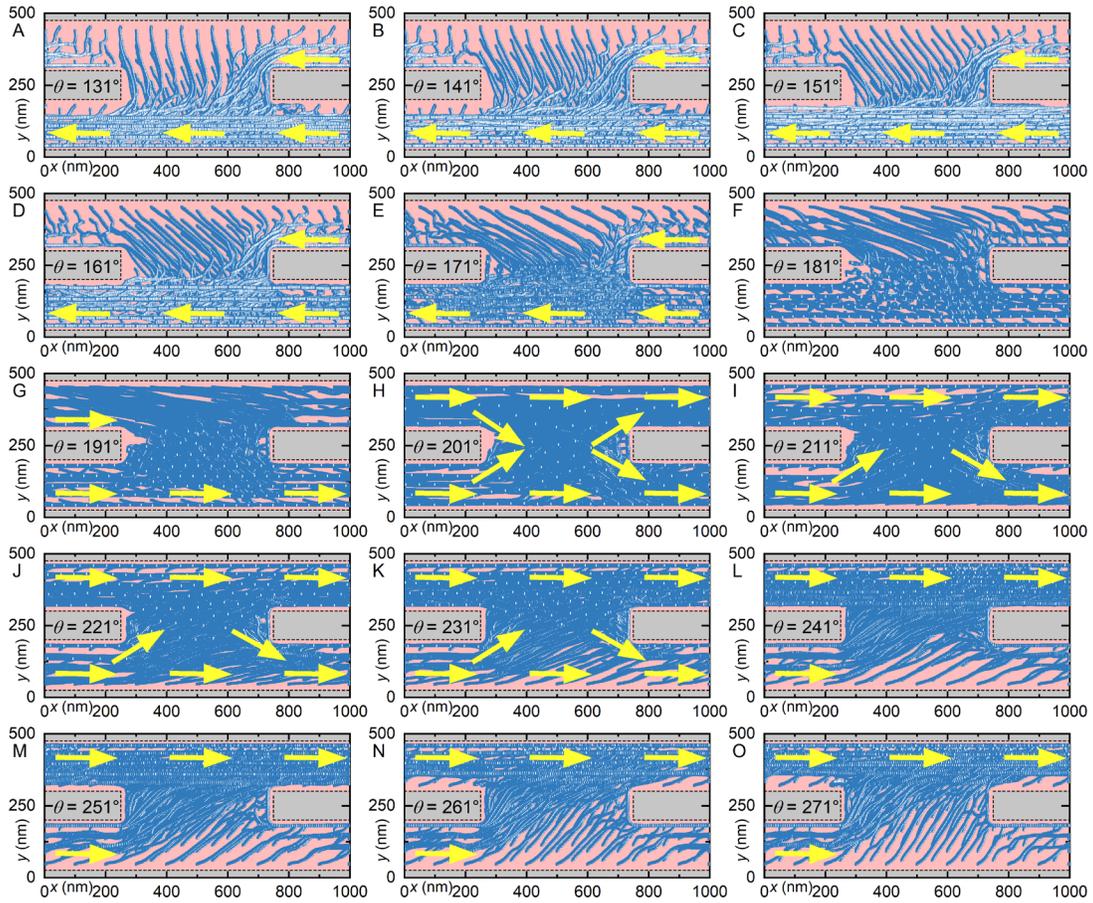

**Fig. 4. Overlay of flowing skyrmions during *t* = 0-1000 ns, showing the transitional pathlines (i.e., trajectories) of skyrmions during the overall driving operation.** The intrinsic skyrmion Hall angle is controlled by setting the spin-polarization angle to (***A***) 131, (***B***) 141, (***C***) 151, (***D***) 161, (***E***) 171, (***F***) 181, (***G***) 191, (***H***) 201, (***I***) 211, (***J***) 221, (***K***) 231, (***L***) 241, (***M***) 251, (***N***) 261, and (***O***) 271 degrees. However, the flow directions of the skyrmions are determined by both the intrinsic skyrmion Hall angle and the skyrmion-skyrmion and skyrmion-edge interactions. The yellow arrows indicate the overall flow directions of the skyrmions in the H-shaped junction.

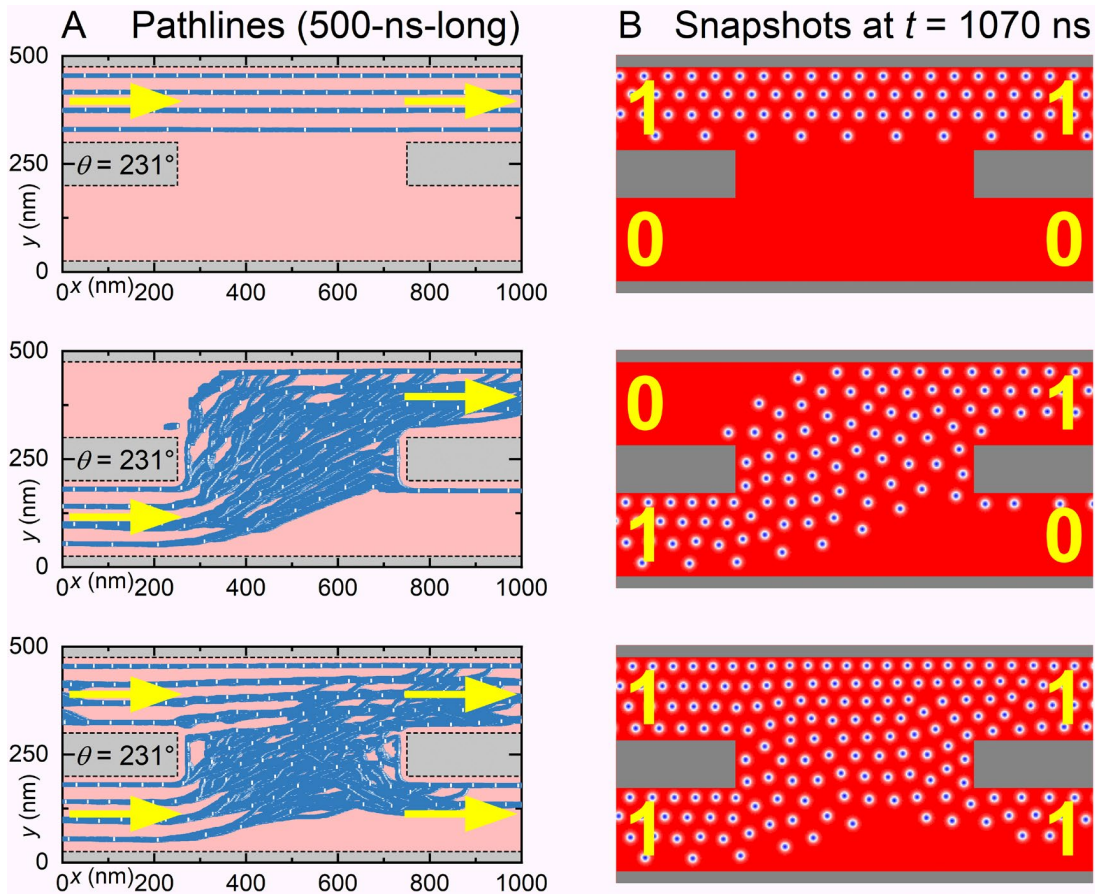

**Fig. 5. Fully developed skyrmion flows that can be used for logical OR and AND operations.**
(***A***) Overlay of flowing skyrmions during $t$ = 500-1000 ns, showing fully developed pathlines of skyrmions. The system is driven by a small current of $j$ = 0.2 MA cm$^{-2}$. The spin polarization angle is fixed at $\theta$ = 231 degrees. The input and output ports (including the vent port) are connected to skyrmion chambers. (***B***) Selected snapshots of the out-of-plane magnetization demonstrating the input and output/vent states in the device. Logical OR and AND gates can be realized depending on the configurations of the MTJ readout sensors (see **Fig. 2*D*** and **Fig. 2*E***).